\documentclass[runningheads]{llncs}
\usepackage[T1]{fontenc}
\usepackage{graphicx}
\usepackage{amsfonts}
\usepackage{amsmath}
\usepackage{algorithm}
\usepackage{algpseudocode}
\usepackage{threeparttable}
\usepackage{booktabs}
\usepackage{multirow}
\usepackage{multicol}
\usepackage{pgfplots}
\usepackage{tabularx}
\usepackage{tikz}
\usepackage{hyperref}
\usepackage{url}

\begin{document}
\title{A Study of Secure Algorithms for Vertical Federated Learning: Take Secure Logistic Regression as an Example\thanks{This paper is accepted by the 20th International Conference on Security \& Management
(SAM 2021). The table of accepted papers can be found in \href{https://www.american-cse.org/static/CSCE21\%20book\%20abstracts.pdf}{https://www.american-cse.org/static/CSCE21\%book\%abstracts.pdf}}}
%
%
\author{Huan-Chih Wang\inst{1}\orcidID{0000-0002-1293-1510} \and
Ja-Ling Wu\inst{1}}
\authorrunning{H. Wang et al.}
%
\institute{National Taiwan University, Taiwan \\
\email{\{whcjimmy,wjl\}@cmlab.csie.ntu.edu.tw}}
\maketitle              
\begin{abstract}
After entering the era of big data, more and more companies build services with machine learning techniques. However, it is costly for companies to collect data and extract helpful handcraft features on their own. Although it is a way to combine with other companies’ data for boosting the model’s performance, this approach may be prohibited by laws. In other words, finding the balance between sharing data with others and keeping data from privacy leakage is a crucial topic worthy of close attention. This paper focuses on distributed data and conducts secure model-training tasks on a vertical federated learning scheme. Here, secure implies that the whole process is executed in the encrypted domain; therefore, the privacy concern is released.

\end{abstract}

\section{Introduction}
With the success and maturity of machine learning techniques, many companies choose to build applications based on big data and learning-based approaches. Nevertheless, to get a well-performed model, collecting enough data and discovering meaningful features are basic data scientists' challenges. Most of the time, it is a crucial issue for companies to conquer these difficulties. One way to solve this problem is to exchange their data with other companies to enrich their dataset. For example, a hospital may want to promote its insurance plans to different classes of patients. However, the hospital only has data about patients' health conditions and has no ideas about theirs income or other financial-related information. Intuitively, the hospital may want to request data from banks to train a risk estimation model more precisely, but this behavior is prohibited by law.

With the rise of privacy consciousness worldwide, many companies set up laws to regulate personal data usage. Undoubtedly, regulations in data usage limit the development of the machine learning field. How to exchange data safely from privacy leakage and boost model performance simultaneously becomes a top-listed research issue. In this work, we model the above hospital’s insurance-enhancing example into a vertical federated learning scheme closer to a more realistic situation in the real world. We use recent booming homomorphic encryption cryptosystems to release privacy concerns to prevent leaking data and learn a model directly in the ciphertext domain. In the above example settings, we learn a regression model with data possessed by two parties, but from the be-ginning to the end, these two parties do not know the data owned by the others.
\section{Related Work}
There are plenty of works referring to how to transfer machine learning algorithms properly to fit privacy requirements. In general, we can conclude these works into the training and testing phases. In the testing phase, we do see many needs for privacy-preserving inferencing data in various fields. For example, \cite{wood2020homomorphic} shows plenty of medical applications based on clients’ data, \cite{kim2019efficient} shows a combination of a privacy-preserving model and a blockchain network, and \cite{li2018privacy} builds a system with the power of cloud computing.

When it comes to the training phase, more and more researchers use homomorphic encryption as a tool to meet privacy needs in recent years. As far as we know, both supervised and unsupervised learning algorithms can be learned in the encrypted domain entirely. For instance, \cite{bonte2018privacy,cheon2017homomorphic,chen2018logistic,kim2018secure,yang2019parallel} train a logistic regression model, \cite{park2020he} and \cite{gonzalez2017training} discuss how to learn an SVM model, \cite{liu2020towards} and \cite{akavia2022privacy} talk about decision tree algorithms, \cite{catak2020practical} tries to train a clustering model and \cite{truex2019hybrid} trains a neural network in the encrypted domain.

After Google presented the first federated learning (FL) framework \cite{mcmahan2017communication}, it attracts lots of data scientists’ eyes because FL provides a way to train a model with distributed data directly. However, \cite{yang2019federated} and \cite{li2020federated} showed that information leakages might occur when using an FL framework to train a mod-el. As a result, there is still a need to use security mechanisms to protect data, which is still a new re-search field.  There are a few researchers who devoted themselves to this topic. For example, \cite{xu2019hybridalpha} and \cite{laur2006cryptographically} use homomorphic encryption schemes to train a logistic regression model in the encrypted domain and apply secure multiparty computation techniques to prevent data leakage.

After reading \cite{bonte2018privacy,cheon2017homomorphic,chen2018logistic,kim2018secure,yang2019parallel}, we found that a logistic regression model cannot separate a nonlinear dataset properly. Besides, as pointed out in \cite{truex2019hybrid}, the associated neural network structure highly depends on the tested dataset. As a result, we think the logistic regression with kernel functions may be a solution to the nonlinearly separable datasets. As far as we know, \cite{laur2006cryptographically} is the only paper discussing learning an SVM model with kernel matrices and few works focus on learning a kernel regression model.
\section{Preliminaries}
\subsection{Vertical Federated Learning}
In the real world, data are prohibited from exchanging between different companies by law. \cite{mcmahan2017communication} presented a privacy-preserving framework – federated learning, to learn a model from decentralized data to break the limitation. As \cite{yang2019federated} mentioned, there are three types of federated learning, and we focus on vertical federated learning in this work only. We assume three involved parties in our simplified vertical federated learning scheme: Alice, Bob, and Eve. In the rest of this work, we adopt the following settings: Alice represents a bank with financial-related information and Bob represents a hospital with patients’ health conditions. Because Bob wants to learn a model, Bob also puts labels on each one of his patients. Eve is a person helping Alice and Bob build and update models.

Generally, Alice’s clients and Bob’s patients are not the same, so they need to run a protocol to find their users' intersection. In this paper, we assume Alice and Bob have already got the intersection that contains $N$ parallel users. Suppose Alice’s data has dimension $d_A$ and Bob’s data has dimension $d_B$, that means Alice has $data_A\in \mathbb{R}^{N\times d_A}$ and Bob owns $data_B\in \mathbb{R}^{N\times d_B}$. When concatenating two data together vertically, it generates an aggregated dataset that is used for training. We denoted the aggregated data as \textbf{data}, that is:
\begin{equation*}
    \textbf{data}=[data_A|data_B].
\end{equation*}

Note that the aggregated dataset $\textbf{data}\in \mathbb{R}^{N\times D}$, where $D=d_A+d_B$.

In this paper, we assume Alice and Bob both are “curious-but-honest,” and they do not want to reveal their plaintext data to each other. With this constraint, we have to design several secure protocols in later sections to keep data safe to learn a model jointly.

\subsection{Homomorphic Encryption Cryptosystem}
Homomorphic encryption is a class of encryption algorithms in which certain operations can be directly carried out on ciphertexts to generate the corresponding encrypted results. When decrypting the encrypted result, it will match the same procedures performed on the corresponding plaintexts \cite{wang2024machinelearningbasedsecureface}. To be more precise, homomorphic encryption is an asymmetric cryptosystem, which contains a public key pk and a secret key sk. We denote $E_{PK}(\cdot)$ as an encryption function with the public key pk and $D_{SK}(\cdot)$ a decryption function with the private key sk. Besides, we also denote $m$ as a message in the plaintext domain and $[m]$ the message in the ciphertext domain. As a result, $[m]=E_{PK}(m)$ and $m=D_{PK}([m])$. As \cite{aslett2015review} shows, if there is an addition operation $+$ in the plaintext domain and there is also a corresponding operation $\oplus$ in the ciphertext domain, then the equation
\begin{equation}
    D_{SK}(E_{PK}(m_1)\oplus E_{PK}(m_2))=m_1+m_2
\end{equation}
tells of the homomorphic property.

There are three kinds of homomorphic encryption cryptosystems: partial homomorphic encryption, somewhat homomorphic encryption, and fully homomorphic encryption. Partial homomorphic encryption only supports additions (e.g., Paillier \cite{paillier1999public}) or multiplications (such as RSA \cite{rivest1978method} or ElGamal \cite{elgamal1985public}, etc.) in the encryption domain. Somewhat homomorphic encryption can calculate both additions and multiplications with limited computations times, and in 2009, Gentry \cite{gentry2009fully} designs an algorithm with a bootstrapping method that calculates additions and multiplications without limits.

In our current study, we apply CKKS homomorphic encryption schemes \cite{cheon2017homomorphic} for practice. CKKS scheme approximates several encrypted operations to accelerate computation times. This scheme can also directly encrypt floating-point numbers into the encrypted domain, which helps execute ma-chine learning tasks directly on the ciphertext domain.

\subsection{Mathematical Theorems of Logistic Regression Models}
Reference \cite{ezukwoke2019logistic} introduces two logistic regression models that are popular in the machine learning field. Logistic regression uses a logistic function to transform the confidence of the model from 0 to 1. Equation (2) shows the definition of a logistic function, $\theta(s)$, and as illustrated by its characteristic curve, $\theta(s)$ is continuous for all $s\in \mathcal{R}$. In other words, $\theta(s)$ is always differentiable for $s\in \mathcal{R}$:
\begin{equation}
    \theta(s)=\frac{e^s}{1+e^s}=\frac{1}{1+e^{-s}}.
\end{equation}

\subsubsection{Logistic Regression}
Considering that there is a dataset with $N$ pairs of samples and labels, which are $(x_1, y_1$, $(x_2, y_2)$, $\cdots$, $(x_N, y_N)$ and each sample contains a d-dimensional feature. With a d-dimensional model $w$, the often-adopted objective function in logistic regression, that is:
\begin{equation}
    \min_w E_{in}(w)=\frac{1}{N}\sum_{n=1}^{N}{\log(1+e^{-y_nw^Tx_n})}.
\end{equation}

During the optimization process, taking the stochastic gradient of the cost function is a must. The stochastic gradient descent of the logistic regression given in (3) can be represented as:
\begin{equation}
\begin{split}
    \nabla E_{in}(w) &= \frac{1}{N}\sum_{n=1}^{N}{\frac{e^{-y_nw^Tx_n}}{1+e^{-y_nw^Tx_n}}(-y_nx_n)} \\ 
                    &= \frac{1}{N}\sum_{n=1}^{N}{\theta(-y_nw^Tx_n)(-y_nx_n)}.
\end{split}
\end{equation}

Because of the computational limitation of nonlinear functions in the encrypted domain, we follow \cite{kim2018secure} to transform the sigmoid function into a polynomial function. That is:
\begin{equation}
    \theta(s)\approx a_0+a_1s+a_2s^2+a_3s^3+O(s^4).
\end{equation}
Substituting (5) into (4), the gradient descent of the logistic regression can be written as: (where we omit the terms with orders higher than three in the polynomial function)
\begin{multline}
    \nabla E_{in}(w) = \\ \frac{1}{N}\sum_{n=1}^{N}{(a_0+a_1(-y_nw^Tx_n)+a_2(-y_nw^Tx_n)^2+a_3(-y_nw^Tx_n)^3)(-y_nx_n)}.
\end{multline}

\subsubsection{Kernel Logistic Regression}
In the family of logistic regression, the kernel logistic regression has the reputation of providing higher precision in classification tasks. Next, let us apply the same technique to explore the kernel logistic regression further. Based on the representer theorem \cite{kimeldorf1971some}, for any $L_2$-regularized linear model, we can prove that Equation (7) has an optimal solution: 
\begin{equation}
\begin{split}
    w^* &= \sum_{n=1}^{N}{\beta_nz_n} \\
        &= \frac{\lambda}{N}w^Tw+\frac{1}{N}\sum_{n=1}^{N}{err(y_n, w^Tz_n)}.
\end{split}
\end{equation}

By the representer theorem, the kernel logistic regression can be rewritten as:
\begin{equation}
    E_{ink}(\beta)=\frac{\lambda}{N}\beta^TK\beta+\frac{1}{N}\sum_{n=1}^{N}{\log(1+e^{-y_n\beta^TK(:,n)})}.
\end{equation}

Note that in Equation(8), $\beta$ and $K(:, n)$ are both $N\times 1$ arrays, and $K(:, n)$ is an array containing $[x_1x_n, x_2x_n, \cdots, x_Nx_n]$. Follow the same derivation of Equation (6) and transform the sigmoid function into a cubic polynomial function, the stochastic gradient descent of the kernel logistic regression becomes:
\begin{equation}
    \nabla E_{ink}(\beta)\approx \frac{2\lambda}{N}K^T\beta+\frac{1}{N}\sum_{n=1}^{N}{\left( \sum_{i=0}^{3}{a_i(-y_n\beta^TK(:, n))^i} \right) (-y_nK(:,n))}.
\end{equation}

\subsubsection{The Kernel Functions}
As \cite{abbasnejad2012survey} described, the kernel function elegantly transfers data to a higher-dimensional space, so it has better performance to solve nonlinear problems than the approaches of artificial neural networks and decision trees. After introducing regression models, we are now raising three different kernels \cite{wikipediaKernelMethod} commonly used in regression-related applications.
\begin{enumerate}
    \item Linear Kernel, where $K_l(x_n, x_m) = x_nx_m$.
    \item Polynomial Kernel, where $K_p(x_n, x_m) = (x_nx_m+c)^{d_{poly}}$.
    \item RBF kernel, where $K_{RBF}(x_n, x_m)=e^{-\gamma ||x_n-x_m||^2}$.
\end{enumerate}

To face the more challenging needs of privacy-preserving computation, we apply the Taylor series expansion to approximate the RBF kernel. After some simple calculations, the second-order Taylor series expansion of the RBF kernel becomes:
\begin{equation}
    K_{RBF}(x_n, x_m)=1+(-\gamma||x_n-x_m||^2)+\frac{(-\gamma||x_n-x_m||^2)}{2}.
\end{equation}
\section{Learning Models in the Encrypted Domain}
\subsection{Data Exchange Protocols}
Here, we will design four data exchange protocols, and three of them are kernel-based protocols associated with each kernel function. Let us define $x$ as a vector and $x_i^A$ the $i$-th row (or basis) of the feature vectors (or space) in the dataset $data_A$. Let $K_{ij}^A$ denote a matrix generated by Alice containing the kernel function transition results associated with the feature vectors $x_i^A$ and $x_j^A$.

Because computations in the encrypted domain are costly and multiplications are more expensive than additions, we will try to design protocols that contain multiplications as little as possible and control multiplication times precisely so that we can get correct results back into the plaintext domain. At the end of each section, we will calculate how much computations it will cost in the encrypted field.

\subsubsection{Secure Data Exchange Protocol}
Bob gets $[x]_E$ at the end of this protocol, and the computational cost for $[x]_E$ is 0.
\begin{algorithm}
    \caption{Secure Data Exchange Protocol}
    \begin{algorithmic}[1]
    \Statex \textbf{Eve}
    \State Eve generates a key pair $(pk_E, sk_E)$ for a chosen homomorphic encryption algorithm.
    \State Eve sends the public key $pk_E$ to Alice and Bob.
    \Statex \textbf{Alice}
    \State Alice encrypts $x_i^A$ with Eve's public key which is $[x_i^A]_E$ and sends them to Bob.
    \Statex \textbf{Bob}
    \State Bob encrypts $x_i^B$ with Eve's public key, yielding $[x_i^B]_E$.
    \State Bob concatenates $[x_i^A]_E$ and $[x_i^B]_E$ with rotation operation presented in \cite{laine2017simple}, and it becomes $[x]_E$.
    \end{algorithmic}
\end{algorithm}

\subsubsection{Secure Linear Kernel Data Exchange Protocol}
Recall that $K_l(x_n,x_m)=x_nx_m$. Hence, a ciphertext domain linear kernel data exchange protocol can be shown as follow. In this protocol, the computational cost is only one addition.
\begin{algorithm}
    \caption{Secure Linear Kernel Protocol}
    \begin{algorithmic}[1]
    \Statex \textbf{Eve}
    \State Eve generates a key pair $(pk_E, sk_E)$ for a chosen homomorphic encryption algorithm.
    \State Eve sends the public key $pk_E$ to Alice and Bob.
    \Statex \textbf{Alice}
    \State Alice computes the dot production of $x_i^A$ and $x_j^A$. That is $K_{ij}^A=\langle x_i^A\cdot x_j^A\rangle=K^A$.
    \State Alice encrypts $K^A$ with Eve's public key which is $[K^A]_E$ and sends it to Bob.
    \Statex \textbf{Bob}
    \State Bob computes $K^B=\langle x_i^B\cdot x_j^B\rangle$.
    \State Bob encrypts $K^B$ with Eve's public key, yielding $[K^B]_E$.
    \State Bob encrypts $x_i^B$ with Eve's public key, yielding $[x_i^B]_E$.
    \State Bob computes $[K_l]_E=[K^A]_E+[K^B]_E.$
    \end{algorithmic}
\end{algorithm}

\subsubsection{Secure Polynomial Kernel Data Exchange Protocol}
Recall that $K_p(x_n,x_m)=(x_nx_m+c)^(d_{poly})$. Hence, a ciphertext domain polynomial kernel data exchange protocol can be shown as follow. In the secure polynomial kernel data exchange protocol, with a polynomial kernel of degree $d_{poly}$, the computational costs are two additions and $d_{poly}-1$ multiplications.
\begin{algorithm}
    \caption{Secure Polynomial Kernel Exchange Protocol}
    \begin{algorithmic}[1]
    \Statex \textbf{Eve}
    \State Eve generates a key pair $(pk_E, sk_E)$ for a chosen homomorphic encryption algorithm.
    \State Eve sends the public key $pk_E$ to Alice and Bob.
    \Statex \textbf{Alice}
    \State Alice computes the dot production of $x_i^A$ and $x_j^A$. That is $K_{ij}^A=\langle x_i^A\cdot x_j^A\rangle=K^A$.
    \State Alice encrypts $K^A$ with Eve's public key which is $[K^A]_E$ and sends it to Bob.
    \Statex \textbf{Bob}
    \State Bob computes $K^B=\langle x_i^B\cdot x_j^B\rangle$.
    \State Bob encrypts $K^B$ with Eve's public key, yielding $[K^B]_E$.
    \State Bob computes $[K_1]_E=[K^A]_E+[K^B]_E$.
    \State Bob computes $[K_2]_E=[K^1]_E+[c]_E$.
    \For
        \State Bob computes $[K_p]=[K_2]_E\times [K_2]_E$.
    \EndFor
    \end{algorithmic}
\end{algorithm}

\subsubsection{Secure RBF Kernel Data Exchange Protocol}
Recall that the second-order Taylor series expansion of the RBF kernel is $K_{RBF}(x_n, x_m)=1+(-\gamma||x_n-x_m||^2)+\frac{(-\gamma||x_n-x_m||^2)}{2}$; therefore, the corresponding ciphertext domain RBF kernel data exchange protocol can be shown as follow. In the secure RBF kernel data exchange protocol, the computational costs are four additions and one multiplication.
\begin{algorithm}
    \caption{Secure RBF Kernel Exchange Protocol}
    \begin{algorithmic}[1]
    \Statex \textbf{Eve}
    \State Eve generates a key pair $(pk_E, sk_E)$ for a chosen homomorphic encryption algorithm.
    \State Eve sends the public key $pk_E$ to Alice and Bob.
    \Statex \textbf{Alice}
    \State Alice computes $K_{ij}^A=-\gamma||x_i^A=x_j^A||^2=K^{A_1}$ and $K^{A_2}=\frac{1}{\sqrt{2}}K^{A_1}$.
    \State Alice encrypts $K^{A_1}$ and $K^{A_2}$ with Eve's public key which results in $[K^{A_1}]_E$ and $[K^{A_2}]_E$.
    \State Alice sends $[K^{A_1}]_E$ and $[K^{A_2}]_E$ to Bob.
    \Statex \textbf{Bob}
    \State Bob computes $K_{ij}^B=-\gamma||x_i^B=x_j^B||^2=K^{B_1}$ and $K^{B_2}=\frac{1}{\sqrt{2}}K^{B_1}$.
    \State Bob encrypts $K^{B_1}$ and $K^{B_2}$ with Eve's public key separately which results in $[K^{B_1}]_E$ and $[K^{B_2}]_E$.
    \State Bob computes $[K_1]_E=[K^{A_1}]_E+[K^{B_1}]_E$ and $[K_2]_E=[K^{A_2}]_E+[K^{B_2}]_E$.
    \State At last, Bob computes $[k_{RBF}]=1+[K_1]_E+[K_2]_E^2$.
    \end{algorithmic}
\end{algorithm}

Table~\ref{tab_comp_cost} summarizes the computational consumptions of every data exchange protocol. This information helps us select a proper polynomial degree to approximate in the next chapter and set suitable encryption parameters in experiments.
\begin{table}[htb]
\centering
\caption{The computational cost for each kernel data exchange protocol.}\label{tab_comp_cost}
\begin{tabular}{lrr}
    \toprule
    Name & Number of Additions & Number of Multiplications \\
    \midrule
    Secure Data Exchange Protocol & 0 & 0 \\
    Secure Linear Kernel Protocol & 1 & 0 \\
    Secure Polynomial Kernel Protocol & 2 & $d_{poly}-1$ \\
    Secure RBF Kernel Protocol & 4 & 1 \\
    \bottomrule
\end{tabular}
\end{table}

\subsection{Depths of Learning Models in the Encrypted Domain}
In a homomorphic encryption scheme, the cost of multiplication is much higher than an addition. Besides, without the batching method applied, the number of multiplications is limited. As a result, we need to evaluate the times of multiplications when training a model carefully. Here we define the term “depth” which means the time of continuous expansions. For example, the depth of $x^3$ is 2 and the depth of $x^5$ is 4. 

This paper cares about multiplications between a ciphertext with a ciphertext and a ciphertext with a plaintext. We want to know the most extended depth of learning a model to select proper encryption parameters. Table~\ref{tab_depth} shows the depths of various regression models based on the different degrees of polynomial-simulated (poly-simulated) sigmoid functions.
\begin{table}[htb]
\centering
\caption{The relationships between the De-gree of adopted poly-simulated sigmoid functions and the total required Depth of regression models.}\label{tab_depth}
\begin{tabularx}{\textwidth}{|X|c|c|c|c|c|c|}
    \toprule
    The degree of poly-simulated sigmoid function & 1 & 2 & 3 & 4 & 5 & $\cdots$ \\ 
    \midrule
    The total depth of LR                         & 3 & 4 & 5 & 5 & 6 & $\cdots$ \\
    \midrule
    The total depth of KLR with linear kernel     & 4 & 5 & 6 & 6 & 7 & $\cdots$ \\
    \midrule
    The total depth of KLR with polynomial kernel & $d_{poly}$+2 & $d_{poly}$+3 & $d_{poly}$+4 & $d_{poly}$+4 & $d_{poly}$+ 5 & $\cdots$ \\
    \midrule
    The total depth of KLR with RBF kernel        & 5 & 6 & 7 & 7 & 8 & $\cdots$ \\
    \bottomrule
\end{tabularx}
\end{table}

\subsection{Secure Model Training Protocol}
\subsubsection{Logistic Regression}
Now, we will build the privacy-preserving protocols for calculating the stochastic gradient de-scent, which is a must-have process in the vertical federated learning scheme.
\begin{algorithm}
    \caption{Secure Logistic Regression Training Protocol}
    \begin{algorithmic}[1]
    \Statex \textbf{Eve}
    \State Eve generates a key pair $(pk_E, sk_E)$ for a chosen homomorphic encryption algorithm.
    \State Eve sends the public key $pk_E$ to Alice and Bob.
    \State Eve generates a model $w$ with $D$ dimension, encrypts this model with her public key and sends it to Bob.
    \Statex \textbf{Alice and Bob}
    \State Alice and Bob run a secure data exchange protocol together and Bob gets $[x]_E$ at last. 
    \Statex \textbf{Bob}
    \State Bob decides the learning rate $\lambda$ and a polynomial of degree $d$ to simulate a sigmoid function with parameters $(a_0, a_1,\cdots, a_{d-1}$.
    \For{$0 \leq i < N$}
        \State Bob calculates $[t]_E=w\cdot [x_i]_E$.
        \State Bob applies the tree method \cite{githubGitHubMarwanNourSEALFYPLogisticRegression} to generate $[t]_E$, $[t^2]_E$, $\cdots$, $[t^{d-1}]_E$.
        \State Bob calculates $\alpha_k=-a_ky_i^{k+1}/N \forall 0 \leq k < d$.
        \State Bob calculates $[T]_E=\sum_{k=0}^{d-1}{\alpha_k\times [t^k]_E\times [x_i]_E}$.
    \EndFor
    \State Bob sends $[T]_E$ and $\lambda$ to Eve.
    \Statex \textbf{Eve}
    \State Eve decrypts $[T]_E$ with her private key.
    \State Eve computes $w=w-\lambda\times T$.
    \Statex \textbf{Eve and Bob}
    \State Bob and Eve iterate steps 6 to 13 several times until $w$ converges.
    \end{algorithmic}
\end{algorithm}

Recall that the polynomial approximation of stochastic gradient descent of logistic regression is shown in Equation (6). The secure model training protocol of a logistic regression model can be delivered as follows.

\subsubsection{Kernel Logistic Regression}
Take the kernel logistic regression model with linear kernel as an example. The protocol for training a kernel logistic regression model becomes:
\subsubsection{Logistic Regression}
Now, we will build the privacy-preserving protocols for calculating the stochastic gradient de-scent, which is a must-have process in the vertical federated learning scheme.
\begin{algorithm}
    \caption{Secure Kernel Logistic Regression Training Protocol}
    \begin{algorithmic}[1]
    \Statex \textbf{Eve}
    \State Eve generates a key pair $(pk_E, sk_E)$ for a chosen homomorphic encryption algorithm.
    \State Eve sends the public key $pk_E$ to Alice and Bob.
    \State Eve generates a model $w$ with $D$ dimension, encrypts this model with her public key and sends it to Bob.
    \Statex \textbf{Alice and Bob}
    \State Alice and Bob run a secure Polynomial/RBF kernel exchange protocol together and Bob gets $[K]_E$ at last. 
    \Statex \textbf{Bob}
    \State Bob decides the learning rate $\lambda$ and a polynomial of degree $d$ to simulate a sigmoid function with parameters $(a_0, a_1,\cdots, a_{d-1}$.
    \For{$0 \leq i < N$}
        \State Bob calculates $[t]_E=w\cdot [K(:,i)]_E$.
        \State Bob applies the tree method \cite{githubGitHubMarwanNourSEALFYPLogisticRegression} to generate $[t]_E$, $[t^2]_E$, $\cdots$, $[t^{d-1}]_E$.
        \State Bob calculates $\alpha_k=-a_ky_i^{k+1}/N \forall 0 \leq k < d$.
        \State Bob calculates $[T]_E=\sum_{k=0}^{d-1}{\alpha_k\times [t^k]_E\times [K(:,i)]_E}$.
    \EndFor
    \State Bob sends $[T]_E$ and $\lambda$ to Eve.
    \Statex \textbf{Eve}
    \State Eve decrypts $[T]_E$ with her private key.
    \State Eve computes $w=w-\lambda\times T$.
    \Statex \textbf{Eve and Bob}
    \State Bob and Eve iterate steps 6 to 13 several times until $w$ converges.
    \end{algorithmic}
\end{algorithm}
\section{Experiments}
First, we want to discover the accuracy with different approximated sigmoid functions and regression functions in this section. After that, we will measure the execution time and RAM usage of learning a model in the encrypted domain. Finally, with the prior knowledge of a specific regression model's computational consumption, we will compare execution time based on various encryption parameters.

We use Microsoft SEAL \cite{laine2017simple} to conduct homomorphic encryption computations in the encrypted do-main and apply OpenMP \cite{openmpHomeOpenMP} to whole parallel processes. All experiments are run on a server with an Intel Xeon E5-2650 v3 CPU and 128 GB RAM.

\subsubsection{Datasets}
To examine the regression models’ performance, we use the make circles \cite{scikitlearnMakecircles} and make moons \cite{scikitlearnMakemoons} datasets presented in Scikit-learn. Make circles is a dataset with a large circle containing a small circle inside, while make moons is a dataset making two interleaving half-circles. Both of them are sitting in a two-dimensional space.  We sample 500 data points from each dataset. Both are nonlinear separable datasets and these datasets are helpful to realize how kernel functions and approximated sigmoid functions affect the accuracy performance.

We divide each dataset into two parts vertically for two parties to apply them into a vertical federated learning scheme. Alice has 500 samples with the first feature and Bob has another 500 samples with the second feature and the associated labels.

\subsubsection{The Accuracy Performance}
To evaluate performances with approximations, we take the performance in the plaintext do-main as a baseline. We use the original sigmoid function and the original RBF kernel in the plaintext domain, and therefore, is no accuracy loss. Table 3 and Figure~\cite{fig1} show the results of the make circles dataset. Also, Table 4 and Figure~\cite{fig2} show the results of the make moons dataset. In Table 3 and Table 4, the first row represents learning a model in the plaintext domain and the rest of the rows represent learning a model in the ciphertext domain. Therefore, in the second and the third rows, we need to use the Taylor series to approximate the RBF kernel.

With kernel functions' help, in the make circles dataset, kernel logistic regression models perform better than the logistic regression model. In the make moons dataset, although KLR with polynomial kernel does not have better accuracy than the logistic regression model, the kernel logistic regression model with RBF kernel is still overkilling other models.

We do experiments with several real-world datasets as well. Table 5 describes the datasets' information, and Table 6 shows our results compared with that of \cite{kim2018secure} and \cite{kim2018logistic}. As Table 6 described, our approximated sigmoid functions perform better in general than \cite{kim2018secure} and \cite{kim2018logistic}. Besides, KLR models perform better than LR models in the Low-Birth-Weight-Study dataset. However, LR and KLR models perform nearly the same in the Prostate-Cancer-Study dataset.
\begin{figure}[htb]
    \centering
    \includegraphics{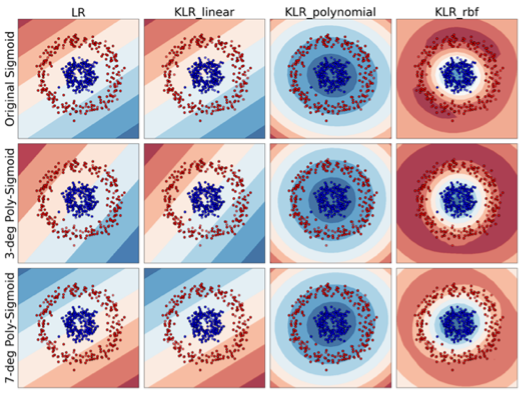}
    \caption{Snapshots of the evolution of Classification processes for the make circles dataset.} 
    \label{fig1}
\end{figure}
\begin{figure}[htb]
    \centering
    \includegraphics{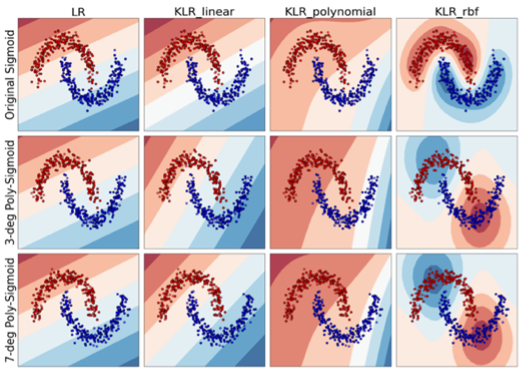}
    \caption{Snapshots of the evolution of the classification processes for the make moons dataset.} 
    \label{fig2}
\end{figure}
\begin{table}[htb]
\centering
\caption{Model Accuracy of the Make circles Dataset.}\label{tab_acc_circle}
\begin{tabular}{lrrrr}
    \toprule
     & LR & KLR & KLR & KLR \\
    \midrule
    Type of Kernel & None & Linear Kernel & Poly-3 Kernel & RBF Kernel \\
    Original Sigmoid & 0.5036 & 0.4976 & 1.00 & 1.00 \\
    3-deg Poly-Sigmoid & 0.5036 & 0.4976 & 1.00 & 0.994 \\
    7-deg Poly-Sigmoid & 0.5036 & 0.4976 & 1.00 & 0.9984 \\
    \bottomrule
\end{tabular}
\end{table}
\begin{table}[htb]
\centering
\caption{Model Accuracy of the Make moons Dataset.}\label{tab_acc_moons}
\begin{tabular}{lrrrr}
    \toprule
     & LR & KLR & KLR & KLR \\
    \midrule
    Type of Kernel & None & Linear Kernel & Poly-3 Kernel & RBF Kernel \\
    Original Sigmoid & 0.882 & 0.8704 & 0.842 & 1.00 \\
    3-deg Poly-Sigmoid & 0.8696 & 0.81 & 0.8016 & 0.928 \\
    7-deg Poly-Sigmoid & 0.876 & 0.8304 & 0.8144 & 0.97 \\
    \bottomrule
\end{tabular}
\end{table}
\begin{table}[htb]
\centering
\caption{Description of the LogisticDx Dataset.}\label{tab_desc_logdx}
\begin{tabular}{lrr}
    \toprule
    & Numoer of Samples & Number of Features \\
    \midrule
    Low Bith Weight Study (LBW) \cite{rdrrLbwBirth} & 189 & 10 \\
    Prosate Cancer Study (PCS) \cite{rdrrPcsProstate} & 379 & 10 \\
    \bottomrule
\end{tabular}
\end{table}
\begin{table}[htb]
\centering
\caption{Model Accuracy of Low-Birth-Weight Study and Prostate-Cancer-Study Dataset.}\label{tab_acc_logdx}
\begin{tabular}{lrrr}
    \toprule
    Model & Def of Poly-Sigmoid & LBW & PCS \\
    \midrule
    \cite{kim2018secure} & \multirow{5}{*}{3-deg Poly-Sigmoid} & 0.693 & 0.689 \\
    LR  & & 0.825 & 0.758 \\
    KLR Linear Kernel & & 0.857 & 0.750 \\
    KLR Poly-3 Kernel & & 0.852 & 0.718 \\
    KLR Taylor-2 RBF Kernel & & 0.852 & 0.708 \\
    \midrule
    \cite{kim2018logistic} & \multirow{5}{*}{5-deg Poly-Sigmoid} & 0.692 & 0.683 \\
    LR  & & 0.825 & 0.755 \\
    KLR Linear Kernel & & 0.857 & 0.747 \\
    KLR Poly-3 Kernel & & 0.878 & 0.726 \\
    KLR Taylor-2 RBF Kernel & & 0.878 & 0.734 \\
    \midrule
    \cite{kim2018secure} & \multirow{5}{*}{7-deg Poly-Sigmoid} & 0.693 & 0.691 \\
    LR  & & 0.831 & 0.755 \\
    KLR Linear Kernel & & 0.847 & 0.747 \\
    KLR Poly-3 Kernel & & 0.894 & 0.734 \\
    KLR Taylor-2 RBF Kernel & & 0.894 & 0.739 \\
    \bottomrule
\end{tabular}
\end{table}

\subsection{Performance of Training a Model in the Encrypted Domain}
For performance measuring, we take the make circles dataset as the benchmarking example to learn a model in the encrypted domain. All models are trained with 20 iterations. Table 7 compares the execution time of training a model in the plaintext domain and in the encrypted domain. We can find out that training a model in the encrypted domain is a huge task, a tradeoff of privacy concerns. Notice that Table 8 also reports the RAM usage results. Because the feature size of data in a KLR model ($N$ dimension) is larger than an LR model ($D$ dimension), we also need more time and RAM usage to train a KLR model.

Finally, when we realize the depth of a model, we can select suitable encryption parameters to speed up. For example, the encryption parameters 8192 and 16384 in Microsoft SEAL can calculate 5 successive multiplications. Therefore, as Table 9 shown, we can use the parameter 8192 instead of 16384 to train a logistic regression model to save time in calculating gradient descents.
\begin{table}[htb]
\centering
\caption{Execution Time of Training Regression Models.}\label{tab_desc_logdx}
\begin{tabular}{lrrrr}
    \toprule
    & LR & KLR & KLR & KLR \\
    \midrule
    Type of Kernel & None & Linear Kernel & Poly-3 Kernel & Taylor-2 RBF Kernel \\
    Plaintext Domain (s) & 0.310 & 0.5 & 0.770 & 0.680 \\
    Encrypted Domain (s) & 4076.4 & 42302.6 & 29638.1 & 35724.7 \\
    \bottomrule
\end{tabular}
\end{table}
\begin{table}[htb]
\centering
\caption{RAM usage of training a model in the Encrypted Domain.}\label{tab_desc_logdx}
\begin{tabular}{lrrrr}
    \toprule
    & LR & KLR & KLR & KLR \\
    \midrule
    Type of Kernel & None & Linear Kernel & Poly-3 Kernel & Taylor-2 RBF Kernel \\
    RAM (GB) & 5.7 & 12.1 & 11.3 & 19.8 \\
    \bottomrule
\end{tabular}
\end{table}
\begin{table}[htb]
\centering
\caption{Execution Time (in seconds) of a LR model with Different Encryption Parameters.}\label{tab_desc_logdx}
\begin{tabular}{lrr}
    \toprule
    Encryption Parameters & 8192 & 16384 \\
    \midrule
    Encode  & 0.172 & 0.399 \\
    Encrypt & 0.231 & 0.597 \\
    Calculate gradient descent & 753.2 & 4075.4 \\
    \bottomrule
\end{tabular}
\end{table}
\section{Conclusions \& Future Work}
The kernel logistic regression model demonstrates a balance of model accuracy and computa-tion costs in the encrypted domain learning tasks. Compared with the logistic regression model, kernel logistic regression models have a better performance on nonlinearly separable datasets. Compared to neural networks, kernel logistic regression models are much easier to learn in the encrypted domain. We believe that the kernel logistic regression models can be successfully applied in many realistic situations.

However, the drawbacks of learning a kernel logistic regression model are apparent. When the number of samples is far more extensive than the feature’s dimension, which means $N\gg D$, the kernel matrix becomes a heavy burden for computations and RAM usage in the encrypted domain. Besides, the execution time for training a model in the ciphertext domain is so long that it is hard for data scientists to search proper hyper-parameters back and forth. Finding parameters efficiently with distributed data and pre-selecting a suitable model for needs are also issues to solve in the future.

\bibliographystyle{splncs04} 
\bibliography{samplepaper} 

\begin{thebibliography}{10}
\providecommand{\url}[1]{\texttt{#1}}
\providecommand{\urlprefix}{URL }
\providecommand{\doi}[1]{https://doi.org/#1}

\bibitem{githubGitHubMarwanNourSEALFYPLogisticRegression}
{G}it{H}ub - {M}arwan{N}our/{S}{E}{A}{L}-{F}{Y}{P}-{L}ogistic-{R}egression:
  {M}y {F}inal {Y}ear {P}roject using {M}icrosoft {S}{E}{A}{L}: {L}ogistic
  {R}egression over {E}ncrypted {D}ata. --- github.com.
  \url{https://github.com/MarwanNour/SEAL-FYP-Logistic-Regression}, [Accessed
  30-10-2024]

\bibitem{openmpHomeOpenMP}
{H}ome - {O}pen{M}{P} --- openmp.org. \url{https://www.openmp.org/ }, [Accessed
  30-10-2024]

\bibitem{wikipediaKernelMethod}
{K}ernel method - {W}ikipedia --- en.wikipedia.org,
  \url{https://en.wikipedia.org/w/index.php?title=Kernel\_method\&oldid=991967840},
  [Accessed 30-10-2024]

\bibitem{rdrrLbwBirth}
lbw: {L}ow {B}irth {W}eight study data in {L}ogistic{D}x: {D}iagnostic {T}ests
  for {M}odels with a {B}inomial {R}esponse --- rdrr.io.
  \url{https://rdrr.io/rforge/LogisticDx/man/lbw.html }, [Accessed 30-10-2024]

\bibitem{scikitlearnMakecircles}
make\_circles --- scikit-learn.org.
  \url{https://scikit-learn.org/1.5/modules/generated/sklearn.datasets.make_circles.html},
  [Accessed 30-10-2024]

\bibitem{scikitlearnMakemoons}
make\_moons --- scikit-learn.org.
  \url{https://scikit-learn.org/stable/modules/generated/sklearn.datasets.make_moons.html},
  [Accessed 30-10-2024]

\bibitem{rdrrPcsProstate}
pcs: {P}rostate {C}ancer {S}tudy data in {L}ogistic{D}x: {D}iagnostic {T}ests
  for {M}odels with a {B}inomial {R}esponse --- rdrr.io.
  \url{https://rdrr.io/rforge/LogisticDx/man/pcs.html}, [Accessed 30-10-2024]

\bibitem{abbasnejad2012survey}
Abbasnejad, M.E., Ramachandram, D., Mandava, R.: A survey of the state of the
  art in learning the kernels. Knowledge and information systems  \textbf{31},
  193--221 (2012)

\bibitem{akavia2022privacy}
Akavia, A., Leibovich, M., Resheff, Y.S., Ron, R., Shahar, M., Vald, M.:
  Privacy-preserving decision trees training and prediction. ACM Transactions
  on Privacy and Security  \textbf{25}(3),  1--30 (2022)

\bibitem{aslett2015review}
Aslett, L.J., Esperan{\c{c}}a, P.M., Holmes, C.C.: A review of homomorphic
  encryption and software tools for encrypted statistical machine learning.
  arXiv preprint arXiv:1508.06574  (2015)

\bibitem{bonte2018privacy}
Bonte, C., Vercauteren, F.: Privacy-preserving logistic regression training.
  BMC medical genomics  \textbf{11},  13--21 (2018)

\bibitem{catak2020practical}
Catak, F.O., Aydin, I., Elezaj, O., Yildirim-Yayilgan, S.: Practical
  implementation of privacy preserving clustering methods using a partially
  homomorphic encryption algorithm. Electronics  \textbf{9}(2), ~229 (2020)

\bibitem{chen2018logistic}
Chen, H., Gilad-Bachrach, R., Han, K., Huang, Z., Jalali, A., Laine, K.,
  Lauter, K.: Logistic regression over encrypted data from fully homomorphic
  encryption. BMC medical genomics  \textbf{11},  3--12 (2018)

\bibitem{cheon2017homomorphic}
Cheon, J.H., Kim, A., Kim, M., Song, Y.: Homomorphic encryption for arithmetic
  of approximate numbers. In: Advances in Cryptology--ASIACRYPT 2017: 23rd
  International Conference on the Theory and Applications of Cryptology and
  Information Security, Hong Kong, China, December 3-7, 2017, Proceedings, Part
  I 23. pp. 409--437. Springer (2017)

\bibitem{elgamal1985public}
ElGamal, T.: A public key cryptosystem and a signature scheme based on discrete
  logarithms. IEEE transactions on information theory  \textbf{31}(4),
  469--472 (1985)

\bibitem{ezukwoke2019logistic}
Ezukwoke, K., Zareian, S.: Logistic regression and kernel logistic regression a
  comparative study of logistic regression and kernel logistic regression for
  binary classification. University Jean Monnet: Saint-Etienne, France  (2019)

\bibitem{gentry2009fully}
Gentry, C.: A fully homomorphic encryption scheme. Stanford university (2009)

\bibitem{gonzalez2017training}
Gonz{\'a}lez-Serrano, F.J., Navia-V{\'a}zquez, {\'A}., Amor-Mart{\'\i}n, A.:
  Training support vector machines with privacy-protected data. Pattern
  Recognition  \textbf{72},  93--107 (2017)

\bibitem{kim2018logistic}
Kim, A., Song, Y., Kim, M., Lee, K., Cheon, J.H.: Logistic regression model
  training based on the approximate homomorphic encryption. BMC medical
  genomics  \textbf{11},  23--31 (2018)

\bibitem{kim2019efficient}
Kim, H., Kim, S.H., Hwang, J.Y., Seo, C.: Efficient privacy-preserving machine
  learning for blockchain network. Ieee Access  \textbf{7},  136481--136495
  (2019)

\bibitem{kim2018secure}
Kim, M., Song, Y., Wang, S., Xia, Y., Jiang, X., et~al.: Secure logistic
  regression based on homomorphic encryption: Design and evaluation. JMIR
  medical informatics  \textbf{6}(2),  e8805 (2018)

\bibitem{kimeldorf1971some}
Kimeldorf, G., Wahba, G.: Some results on tchebycheffian spline functions.
  Journal of mathematical analysis and applications  \textbf{33}(1),  82--95
  (1971)

\bibitem{laine2017simple}
Laine, K.: Simple encrypted arithmetic library 2.3.1. Microsoft Research
  https://www. microsoft.
  com/en-us/research/uploads/prod/2017/11/sealmanual-2-3-1. pdf  (2017)

\bibitem{laur2006cryptographically}
Laur, S., Lipmaa, H., Mielik{\"a}inen, T.: Cryptographically private support
  vector machines. In: Proceedings of the 12th ACM SIGKDD international
  conference on Knowledge discovery and data mining. pp. 618--624 (2006)

\bibitem{li2018privacy}
Li, P., Li, J., Huang, Z., Gao, C.Z., Chen, W.B., Chen, K.: Privacy-preserving
  outsourced classification in cloud computing. Cluster Computing  \textbf{21},
   277--286 (2018)

\bibitem{li2020federated}
Li, T., Sahu, A.K., Talwalkar, A., Smith, V.: Federated learning: Challenges,
  methods, and future directions. IEEE signal processing magazine
  \textbf{37}(3),  50--60 (2020)

\bibitem{liu2020towards}
Liu, L., Chen, R., Liu, X., Su, J., Qiao, L.: Towards practical
  privacy-preserving decision tree training and evaluation in the cloud. IEEE
  Transactions on Information Forensics and Security  \textbf{15},  2914--2929
  (2020)

\bibitem{mcmahan2017communication}
McMahan, B., Moore, E., Ramage, D., Hampson, S., y~Arcas, B.A.:
  Communication-efficient learning of deep networks from decentralized data.
  In: Artificial intelligence and statistics. pp. 1273--1282. PMLR (2017)

\bibitem{paillier1999public}
Paillier, P.: Public-key cryptosystems based on composite degree residuosity
  classes. In: International conference on the theory and applications of
  cryptographic techniques. pp. 223--238. Springer (1999)

\bibitem{park2020he}
Park, S., Byun, J., Lee, J., Cheon, J.H., Lee, J.: He-friendly algorithm for
  privacy-preserving svm training. IEEE Access  \textbf{8},  57414--57425
  (2020)

\bibitem{rivest1978method}
Rivest, R.L., Shamir, A., Adleman, L.: A method for obtaining digital
  signatures and public-key cryptosystems. Communications of the ACM
  \textbf{21}(2),  120--126 (1978)

\bibitem{truex2019hybrid}
Truex, S., Baracaldo, N., Anwar, A., Steinke, T., Ludwig, H., Zhang, R., Zhou,
  Y.: A hybrid approach to privacy-preserving federated learning. In:
  Proceedings of the 12th ACM workshop on artificial intelligence and security.
  pp. 1--11 (2019)

\bibitem{wang2024machinelearningbasedsecureface}
Wang, H.C.: A machine learning-based secure face verification scheme and its
  applications to digital surveillance. Master's Thesis  \textbf{2017},  1--42
  (2017)

\bibitem{wood2020homomorphic}
Wood, A., Najarian, K., Kahrobaei, D.: Homomorphic encryption for machine
  learning in medicine and bioinformatics. ACM Computing Surveys (CSUR)
  \textbf{53}(4),  1--35 (2020)

\bibitem{xu2019hybridalpha}
Xu, R., Baracaldo, N., Zhou, Y., Anwar, A., Ludwig, H.: Hybridalpha: An
  efficient approach for privacy-preserving federated learning. In: Proceedings
  of the 12th ACM workshop on artificial intelligence and security. pp. 13--23
  (2019)

\bibitem{yang2019federated}
Yang, Q., Liu, Y., Chen, T., Tong, Y.: Federated machine learning: Concept and
  applications. ACM Transactions on Intelligent Systems and Technology (TIST)
  \textbf{10}(2),  1--19 (2019)

\bibitem{yang2019parallel}
Yang, S., Ren, B., Zhou, X., Liu, L.: Parallel distributed logistic regression
  for vertical federated learning without third-party coordinator. arXiv
  preprint arXiv:1911.09824  (2019)

\end{thebibliography}

\end{document}